\documentclass[runningheads, toc]{CMSIM}

\usepackage{graphicx} 
\usepackage[usenames,dvipsnames]{color}

\usepackage{amsmath,amssymb}
\usepackage[retainorgcmds]{IEEEtrantools}

\usepackage[]{mciteplus}
\mciteSetBstSublistMode{n}
\mciteSetMidEndSepPunct{\quad$\bullet$\quad}{.}{\relax}
\usepackage[colorlinks=true, linktoc=page, citecolor=blue, urlcolor=blue]{hyperref}

\renewcommand{\thefootnote}

\title*{\color{Blue}\textbf{M-theory as a dynamical system generator}}

\titlerunning{\it M-theory as a dynamical system}

\author{M. Axenides,\inst{1} E. Floratos,\inst{1,2} D. Katsinis,\inst{1} G. Linardopoulos \inst{1}}

\authorrunning{\it Axenides et al}

\institute{Institute of Nuclear and Particle Physics, N.C.S.R.\ "Demokritos", \\
153 10, Agia Paraskevi, Greece \\[6pt]
\and
Department of Nuclear and Particle Physics, National and Kapodistrian University of Athens, 157 84, Athens, Greece. \\[6pt]
\hangindent=1.3cm E-mails: {\tt axenides@inp.demokritos.gr}, {\tt mflorato@phys.uoa.gr}, {\tt dkatsinis@phys.uoa.gr}, {\tt glinard@inp.demokritos.gr}.}

\begin{document}
\thispagestyle{empty}
\maketitle
\setlength{\leftskip}{0pt}
\setlength{\headsep}{16pt}
\footnote{\begin{tabular}{p{11.2cm}r}
\small {\it $13^{th}$CHAOS Conference Proceedings, 9 - 12 June 2020, Florence, Italy} \\
\small C. H. Skiadas (Ed)\\
\small \textcopyright {} 2020 ISAST & \includegraphics[scale=0.38]{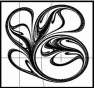}
\end{tabular}}
\vspace{-.5cm} \begin{abstract}
We review our recent work on ellipsoidal M2-brane solutions in the large-$N$ limit of the BMN matrix model. These bosonic finite-energy membranes live inside $\mathfrak{so}\left(3\right) \times \mathfrak{so}\left(6\right)$ symmetric plane-wave spacetimes and correspond to local extrema of the energy functional. They are static in $\mathfrak{so}\left(3\right)$ and stationary in $\mathfrak{so}\left(6\right)$. Chaos appears at the level of radial stability analysis through the explicitly derived spectrum of eigenvalues. The angular perturbation analysis is suggestive of the presence of weak turbulence instabilities that propagate from low to high orders in perturbation theory. \\
\end{abstract}

\section[Introduction]{Introduction}
\setcounter{footnote}{0} \renewcommand{\thefootnote}{\arabic{footnote}}
\paragraph{M-theory} By the end of the first superstring revolution (1984-1994), five seemingly different 10-dimensional superstring theories had emerged:
\begin{center} Types \textrm{I}, \textrm{II} (\textrm{II}A, \textrm{II}B), Heterotic ($\mathfrak{so}(32)$, $\text{E}_8 \times \text{E}_8$). \\[6pt] \end{center}
During the subsequent second superstring revolution (1994-2003), it was found that the 5 superstring theories are connected via a web of dualities (T-duality, S-duality, U-duality, mirror symmetry). What is more, it was realized that the five 10-dimensional superstring theories were just limiting cases of an 11-dimensional theory. This theory was called "M-theory"; it is obtained in the strong-coupling limit ($g_s \rightarrow \infty$) of \textrm{II}A superstring theory. The letter "M" stands for "magic, mystery and matrix" according to one of its founders, E.\ Witten \cite{Witten98m}. Others have associated the letter "M" with "membranes" \cite{Duff04b}.
\paragraph{Relativistic membranes} The idea behind the theory of relativistic membranes is simple: replace 1-dimensional lines (strings) with 2-dimensional surfaces (membranes), much like lines/strings replace 0-dimensional points/particles in the passage from quantum field theory to string theory. Like point particles and strings, membranes are Poincar\'{e} invariant objects that can be supersymmetrized. It has been proven that supermembranes can only be defined consistently in 11 spacetime dimensions. Higher-dimensional extended supersymmetric objects (Mp-branes) can be defined in an analogous fashion. Nonetheless, there are reasons to believe that supermembranes (or "M2-branes") are the fundamental objects of the 11-dimensional M-theory, just like strings are the fundamental objects of 10-dimensional string theory.
\paragraph{Matrix models} According to the \emph{matrix theory conjecture} of Banks, Fischler, Shenker and Susskind (BFSS) \cite{BFSS97}, a theory of matrix-discretized supermembranes provides a realization of M-theory in flat spacetime. In the language of matrix models, membranes are fuzzy objects that are represented by $N\times N$ matrices. In the limit of very large matrix dimensions ($N\rightarrow\infty$), these matrix models are known to reduce to supermembrane theories. \\
\indent In 2002, Berenstein, Maldacena and Nastase (BMN) \cite{BMN02} proposed a reformulation of the BFSS matrix model on a particular type of a background that consists of a weakly curved spacetime that is known as a plane-wave, supported by a constant (4-form) field strength:
\begin{IEEEeqnarray}{ll}
ds^2 = &-2 dx^{+} dx^{-} - \left[\frac{\mu^2}{9}\sum_{i=1}^3 x_i x_i + \frac{\mu^2}{36}\sum_{j=1}^6 y_j y_j\right] dx^+ dx^+ + \sum_{i=1}^3 dx_i dx_i + \nonumber \\
& + \sum_{j=1}^6 dy_j dy_j, \qquad F_{123+} = \mu. \label{MaximallySupersymmetricBackground}
\end{IEEEeqnarray}
Briefly, the BMN matrix model is a deformation of the BFSS matrix model by mass terms and a flux (aka Myers) term. In the large-$N$ limit it is again known \cite{DasguptaJabbariRaamsdonk02} that the BMN matrix model reduces to a theory of supermembranes in the 11-dimensional plane-wave background \eqref{MaximallySupersymmetricBackground}. Interestingly, M(atrix) theory has quite recently been applied to the study of chaotic phenomena that take place on the horizons of black holes.
\paragraph{Black holes} Black holes (BHs) are regions of spacetime where the force of gravity is so strong that nothing (not even light) can escape. The 2-dimensional surface beyond which it is (classically) impossible for matter or information to escape the gravitational pull of a BH is known as the BH's event horizon. In 1974 Stephen Hawking predicted that it is (quantum-mechanically) possible for BHs to emit thermal radiation and thus slowly evaporate. Because Hawking's radiation is purely thermal, all the information that is stored in BHs seems to get lost. \\
\indent To resolve the ensuing BH \emph{information paradox} we ultimately need to understand the mechanisms with which information is being stored and processed in BHs. One such mechanism is known as \emph{fast scrambling} or ultra-fast thermalization \cite{SekinoSusskind08}. More generally, it is widely believed that chaotic phenomena are a dominant feature of BH horizons. Because it is inherently nonlocal, M(atrix) theory turns out to be a valuable tool in the study of information processing by BHs. More precisely, M(atrix) theory can be used to model the dynamics of the microscopic degrees of freedom that are present on BH horizons \cite{Gur-AriHanadaShenker15, AsanoKawaiYoshida15}.
\section[General setup]{General setup}
\noindent Our starting point is the Hamiltonian of a bosonic relativistic membrane in the 11-dimensional maximally supersymmetric plane-wave background \eqref{MaximallySupersymmetricBackground}. The Hamiltonian reads, in the so-called light-cone gauge $x^+ = \tau$ \cite{DasguptaJabbariRaamsdonk02}:
\begin{IEEEeqnarray}{ll}
H = \frac{T}{2}\int d^2\sigma\bigg[&\pi_i^2 + \frac{1}{2}\left\{x_i,x_j\right\}^2 + \frac{1}{2}\left\{y_i,y_j\right\}^2 + \left\{x_i,y_j\right\}^2 + \frac{\mu^2 x^2}{9} + \frac{\mu^2 y^2}{36} - \nonumber \\
& - \frac{\mu}{3}\,\epsilon_{ijk}\left\{x_i,x_j\right\}x_k\bigg]. \qquad \label{ppWaveHamiltonian}
\end{IEEEeqnarray}
From now on the indices of the coordinates $x_i$ will implicitly be taken to run from 1 to 3, while those of the coordinates $y_j$ will run from 1 to 6.\footnote{Note also that there's no distinction between upper and lower indices, so that these will be henceforth used interchangeably.} In \eqref{ppWaveHamiltonian} $T$ stands for the membrane tension and
\begin{IEEEeqnarray}{ll}
\pi_i^2 \equiv \sum_{i=1}^3 \dot{x}_i \dot{x}_i + \sum_{j=1}^6 \dot{y}_j \dot{y}_j, \qquad x^2 \equiv \sum_{i=1}^3 x_i x_i, \qquad y^2 \equiv \sum_{j=1}^6 y_j y_j.
\end{IEEEeqnarray}
The definition of the Poisson bracket $\{f,g\}$ that we will be using is
\begin{equation}
\left\{f\,,\,g\right\} \equiv \frac{\epsilon_{rs}}{\sqrt{w\left(\boldsymbol\sigma\right)}} \, \partial_r f \; \partial_s g = \frac{1}{\sqrt{w\left(\boldsymbol\sigma\right)}} \, \left(\partial_1 f \; \partial_2 g - \partial_2 f \; \partial_1 g\right), \label{PoissonBracket}
\end{equation}
where $d^2\sigma = \sqrt{w\left(\boldsymbol\sigma\right)} \; d\sigma_1 \, d\sigma_2$ is the spatial volume element of the worldvolume and $\epsilon_{rs}$ is the 2-dimensional Levi-Civita symbol. In a flat worldvolume it's $w\left(\boldsymbol\sigma\right) = 1$ and the usual definition of the Poisson bracket is retrieved. \\
\indent The Lagrangian equations of motion for the spatial coordinates $x$ and $y$ corresponding to the Hamiltonian \eqref{ppWaveHamiltonian} are:
\begin{IEEEeqnarray}{l}
\ddot{x}_i = \left\{\left\{x_i,x_j\right\},x_j\right\} + \left\{\left\{x_i,y_j\right\},y_j\right\} - \frac{\mu^2}{9}\,x_i + \frac{\mu}{2}\epsilon_{ijk}\left\{x_j,x_k\right\} \label{xEquation} \\
\ddot{y}_i = \left\{\left\{y_i,y_j\right\},y_j\right\} + \left\{\left\{y_i,x_j\right\},x_j\right\} - \frac{\mu^2}{36}\,y_i. \label{yEquation}
\end{IEEEeqnarray}
The coordinates $x$ and $y$ can also be shown to obey the Gauss law constraint:
\begin{IEEEeqnarray}{l}
\sum_{i = 1}^3\left\{\dot{x}_i, x_i\right\} + \sum_{j = 1}^6\left\{\dot{y}_j, y_j\right\} = 0. \label{GaussLaw1}
\end{IEEEeqnarray}
\section[The spherical ansatz]{The spherical ansatz}
Let us make the following ansatz for the spatial coordinates $x$ and $y$ \cite{CollinsTucker76, HarmarkSavvidy00}:
\begin{IEEEeqnarray}{lll}
x_i \equiv x_{1i} = \tilde{x}_{1i}\left(\tau\right)e_1\left(\sigma\right), \quad & i = 1,\ldots,q_1 \label{SphericalAnsatz1x}\\
x_{q_1+j} \equiv x_{2j} = \tilde{x}_{2j}\left(\tau\right)e_2\left(\sigma\right), \quad & j = 1,\ldots,q_2 \qquad \& \quad & q_1 + q_2 + q_3 = 3 \qquad \label{SphericalAnsatz2x} \\
x_{q_1+q_2+k} \equiv x_{3k} = \tilde{x}_{3k}\left(\tau\right)e_3\left(\sigma\right), \quad & k = 1,\ldots,q_3 \label{SphericalAnsatz3x}
\end{IEEEeqnarray}
and
\begin{IEEEeqnarray}{lll}
y_i \equiv y_{1i} = \tilde{y}_{1i}\left(\tau\right)e_1\left(\sigma\right), \quad & i = 1,\ldots,s_1 \label{SphericalAnsatz1y}\\
y_{s_1+j} \equiv y_{2j} = \tilde{y}_{2j}\left(\tau\right)e_2\left(\sigma\right), \quad & j = 1,\ldots,s_2 \qquad \& \quad & s_1 + s_2 + s_3 = 6 \qquad \label{SphericalAnsatz2y} \\
y_{s_1+s_2+k} \equiv y_{3k} = \tilde{y}_{3k}\left(\tau\right)e_3\left(\sigma\right), \quad & k = 1,\ldots,s_3. \label{SphericalAnsatz3y}
\end{IEEEeqnarray}
The ansatz \eqref{SphericalAnsatz1x}--\eqref{SphericalAnsatz3y} splits the coordinates $x$ and $y$ into three groups
\begin{IEEEeqnarray}{lllll}
x_{ai} = \tilde{x}_{ai}\left(\tau\right) e_a \qquad \& \qquad y_{bj} = \tilde{y}_{bj}\left(\tau\right) e_b,
\end{IEEEeqnarray}
where $i = 1,\ldots,q_a, \ j = 1,\ldots,s_b, \ a,b = 1,2,3$. Going over to spherical coordinates, $\left(\sigma_1, \sigma_2\right) \rightarrow \left(\theta, \phi\right)$, we define:\footnote{We use the volume element in $(\theta,\phi)$ space which implies that $\sqrt{w\left(\boldsymbol\sigma\right)} = \sin\theta$ should be used in the definition \eqref{PoissonBracket} of the Poisson bracket. For alternative parametrizations such as
\begin{IEEEeqnarray}{c}
(e_1, e_2, e_3) = (cn\left(\phi|m\right) sn\left(\theta|n\right), sn\left(\phi|m\right)sn\left(\theta|n\right), sn\left(\theta|n\right)), \label{Epsilon3}
\end{IEEEeqnarray}
where $\phi \in [0,4\mathbb{K}\left(m\right))$ and $\theta \in [0,2\mathbb{K}\left(n\right)]$, the corresponding volume element is $\sqrt{w\left(\boldsymbol\sigma\right)} = sn\left(\theta|n\right) dn\left(\theta|n\right)dn\left(\phi|m\right)$.}
\begin{IEEEeqnarray}{c}
(e_1, e_2, e_3) = (\cos\phi \sin\theta, \sin\phi \sin\theta, \cos\theta), \qquad \phi \in [0,2\pi), \quad \theta \in [0,\pi] \qquad \label{Epsilon1} \\
\{e_i, e_j\} = \epsilon_{ijk} \, e_k, \qquad \int e_i \, e_j\,d^2\sigma = \frac{4\pi}{3} \, \delta_{ij}. \qquad \label{Epsilon2}
\end{IEEEeqnarray}
Note that the Gauss law constraint \eqref{GaussLaw1} is automatically satisfied by the ansatz \eqref{SphericalAnsatz1x}--\eqref{SphericalAnsatz3y}. Now consider the following solutions:
\begin{IEEEeqnarray}{lll}
\tilde{\textbf{x}}_1\left(\tau\right) = e^{\Omega_{x1} \tau} \cdot \tilde{\textbf{x}}_{10}, \qquad & \tilde{\textbf{x}}_2\left(\tau\right) = e^{\Omega_{x2} \tau} \cdot \tilde{\textbf{x}}_{20}, \qquad & \tilde{\textbf{x}}_3\left(\tau\right) = e^{\Omega_{x3} \tau} \cdot \tilde{\textbf{x}}_{30} \qquad \label{Ansatz1} \\
\tilde{\textbf{y}}_1\left(\tau\right) = e^{\Omega_{y1} \tau} \cdot \tilde{\textbf{y}}_{10}, \qquad & \tilde{\textbf{y}}_2\left(\tau\right) = e^{\Omega_{y2} \tau} \cdot \tilde{\textbf{y}}_{20}, \qquad & \tilde{\textbf{y}}_3\left(\tau\right) = e^{\Omega_{y3} \tau} \cdot \tilde{\textbf{y}}_{30}. \qquad \label{Ansatz2}
\end{IEEEeqnarray}
As in the case of flat space (worked out in \cite{AxenidesFloratos07}) it can be shown that the radii
\begin{IEEEeqnarray}{lll}
r_{x1}^2 \equiv \tilde{x}_1^2 = \sum_{i=1}^{q_1} \tilde{x}_{10i} \tilde{x}_{10i}, \ & r_{x2}^2 \equiv \tilde{x}_2^2 = \sum_{j=1}^{q_2} \tilde{x}_{20j} \tilde{x}_{20j}, \ & r_{x3}^2 \equiv \tilde{x}_3^2 = \sum_{k=1}^{q_3} \tilde{x}_{30k} \tilde{x}_{30k} \qquad \label{polar1} \\
r_{y1}^2 \equiv \tilde{y}_1^2 = \sum_{i=1}^{s_1} \tilde{y}_{10i} \tilde{y}_{10i}, \ & r_{y2}^2 \equiv \tilde{y}_2^2 = \sum_{j=1}^{s_2} \tilde{y}_{20j} \tilde{y}_{20j}, \ & r_{y3}^2 \equiv \tilde{y}_3^2 = \sum_{k=1}^{2_3} \tilde{y}_{30k} \tilde{y}_{30k} \qquad \label{polar2}
\end{IEEEeqnarray}
of the ansatz \eqref{Ansatz1}--\eqref{Ansatz2} can be determined (for all the antisymmetric matrices $\Omega_{x1}$, $\Omega_{x2}$, $\Omega_{x3}$, $\Omega_{y1}$, $\Omega_{y2}$, $\Omega_{y3}$) in terms of the conserved angular momenta
\begin{IEEEeqnarray}{ll}
\left(\ell_{x1}\right)_{ij} \equiv \dot{\tilde{x}}_{1i} \tilde{x}_{1j} - \tilde{x}_{1i} \dot{\tilde{x}}_{1j}, \qquad &\left(\ell_{y1}\right)_{ij} \equiv \dot{\tilde{y}}_{1i} \tilde{y}_{1j} - \tilde{y}_{1i} \dot{\tilde{y}}_{1j} \qquad \label{polar3} \\
\left(\ell_{x2}\right)_{ij} \equiv \dot{\tilde{x}}_{2i} \tilde{x}_{2j} - \tilde{x}_{2i} \dot{\tilde{x}}_{2j}, \qquad &\left(\ell_{y2}\right)_{ij} \equiv \dot{\tilde{y}}_{2i} \tilde{y}_{2j} - \tilde{y}_{2i} \dot{\tilde{y}}_{2j} \qquad \label{polar4} \\
\left(\ell_{x3}\right)_{ij} \equiv \dot{\tilde{x}}_{3i} \tilde{x}_{3j} - \tilde{x}_{3i} \dot{\tilde{x}}_{3j}, \qquad &\left(\ell_{y3}\right)_{ij} \equiv \dot{\tilde{y}}_{3i} \tilde{y}_{3j} - \tilde{y}_{3i} \dot{\tilde{y}}_{3j}, \qquad \label{polar5}
\end{IEEEeqnarray}
by minimizing the corresponding effective potential of the membrane. This is completely equivalent to plugging the ansatz \eqref{Ansatz1}--\eqref{Ansatz2} into the equations of motion \eqref{xEquation}--\eqref{yEquation} and determining the relation between the radii $r_{x1}$, $r_{x2}$, $r_{x3}$, $r_{y1}$, $r_{y2}$, $r_{y3}$ and the components of the matrices $\Omega_{x1}$, $\Omega_{x2}$, $\Omega_{x3}$, $\Omega_{y1}$, $\Omega_{y2}$, $\Omega_{y3}$ (which in turn always combine to form the conserved angular momenta $\ell_{x1}$, $\ell_{x2}$, $\ell_{x3}$, $\ell_{y1}$, $\ell_{y2}$, $\ell_{y3}$).
\section[Effective potentials]{Effective potentials}
The energy of the membrane \eqref{ppWaveHamiltonian} becomes:
\begin{IEEEeqnarray}{ll}
E = \frac{2\pi T}{3} \Bigg[&\dot{\tilde{x}}_1^2 + \dot{\tilde{x}}_2^2 + \dot{\tilde{x}}_3^2 + \dot{\tilde{y}}_1^2 + \dot{\tilde{y}}_2^2 + \dot{\tilde{y}}_3^2 + \tilde{x}_1^2 \tilde{x}_2^2 + \tilde{x}_2^2 \tilde{x}_3^2 + \tilde{x}_3^2 \tilde{x}_1^2 + \tilde{y}_1^2 \tilde{y}_2^2 + \tilde{y}_2^2 \tilde{y}_3^2 + \nonumber \\
& + \tilde{y}_3^2 \tilde{y}_1^2 + \tilde{x}_1^2 \left(\tilde{y}_2^2 + \tilde{y}_3^2\right) + \tilde{x}_2^2 \left(\tilde{y}_3^2 + \tilde{y}_1^2\right) + \tilde{x}_3^2 \left(\tilde{y}_1^2 + \tilde{y}_2^2\right) + \frac{\mu^2}{9}\,\tilde{x}^2 + \nonumber \\
& + \frac{\mu^2}{36}\,\tilde{y}^2 - 2\mu\,\epsilon_{ijk}\,\tilde{x}_{1i}\tilde{x}_{2j}\tilde{x}_{3k} \Bigg]. \label{EnergySpherical1}
\end{IEEEeqnarray}
We now proceed to the following decomposition of the coordinates:
\begin{IEEEeqnarray}{ll}
\dot{\tilde{x}}_1^2 \equiv \dot{\tilde{x}}_{1i}\dot{\tilde{x}}_{1i} = \dot{r}_{x1}^2 + \frac{\ell_{x1}^2}{r_{x1}^2}, \qquad \dot{\tilde{y}}_{1}^2 \equiv \dot{\tilde{y}}_{1j}\dot{\tilde{y}}_{1j} = \dot{r}_{y1}^2 + \frac{\ell_{y1}^2}{r_{y1}^2} \label{polar6} \\
\dot{\tilde{x}}_{2}^2 \equiv \dot{\tilde{x}}_{2i}\dot{\tilde{x}}_{2i} = \dot{r}_{x2}^2 + \frac{\ell_{x2}^2}{r_{x2}^2}, \qquad \dot{\tilde{y}}_{2}^2 \equiv \dot{\tilde{y}}_{2j}\dot{\tilde{y}}_{2j} = \dot{r}_{y2}^2 + \frac{\ell_{y2}^2}{r_{y2}^2} \label{polar7} \\
\dot{\tilde{x}}_{3}^2 \equiv \dot{\tilde{x}}_{3i}\dot{\tilde{x}}_{3i} = \dot{r}_{x3}^2 + \frac{\ell_{x3}^2}{r_{x3}^2}, \qquad \dot{\tilde{y}}_{3}^2 \equiv \dot{\tilde{y}}_{3j}\dot{\tilde{y}}_{3j} = \dot{r}_{y3}^2 + \frac{\ell_{y3}^2}{r_{y3}^2}. \label{polar8}
\end{IEEEeqnarray}
Plugging \eqref{polar1}--\eqref{polar2} and \eqref{polar6}--\eqref{polar8} into \eqref{EnergySpherical1}, we find that the energy of the membrane becomes
\begin{IEEEeqnarray}{ll}
E = \frac{2\pi T}{3} \Bigg[&\dot{r}_{x1}^2 + \dot{r}_{x2}^2 + \dot{r}_{x3}^2 + \dot{r}_{y1}^2 + \dot{r}_{y2}^2 + \dot{r}_{y3}^2 + \frac{\ell_{x1}^2}{r_{x1}^2} + \frac{\ell_{x2}^2}{r_{x2}^2} + \frac{\ell_{x3}^2}{r_{x3}^2} + \frac{\ell_{y1}^2}{r_{y1}^2} + \frac{\ell_{y2}^2}{r_{y2}^2} + \nonumber \\
& + \frac{\ell_{y3}^2}{r_{y3}^2} + r_{x1}^2 r_{x2}^2 + r_{x2}^2 r_{x3}^2 + r_{x3}^2 r_{x1}^2 + r_{y1}^2 r_{y2}^2 + r_{y2}^2 r_{y3}^2 + r_{y3}^2 r_{y1}^2 + \nonumber \\
& + r_{x1}^2 \left(r_{y2}^2 + r_{y3}^2\right) + r_{x2}^2 \left(r_{y3}^2 + r_{y1}^2\right) + r_{x3}^2 \left(r_{y1}^2 + r_{y2}^2\right) + \frac{\mu^2}{9}(r_{x1}^2 + \nonumber \\
& + r_{x2}^2 + r_{x3}^2) + \frac{\mu^2}{36}\left(r_{y1}^2 + r_{y2}^2 + r_{y3}^2\right) - 2\mu\,\epsilon_{ijk}\tilde{x}_{1i}\tilde{x}_{2j}\tilde{x}_{3k} \Bigg], \qquad \label{EnergySpherical2}
\end{IEEEeqnarray}
so that the corresponding effective potential reads
\begin{IEEEeqnarray}{ll}
V_{\text{eff}} = \frac{2\pi T}{3} \Bigg[&\frac{\ell_{x1}^2}{r_{x1}^2} + \frac{\ell_{x2}^2}{r_{x2}^2} + \frac{\ell_{x3}^2}{r_{x3}^2} + \frac{\ell_{y1}^2}{r_{y1}^2} + \frac{\ell_{y2}^2}{r_{y2}^2} + \frac{\ell_{y3}^2}{r_{y3}^2} + r_{x1}^2 r_{x2}^2 + r_{x2}^2 r_{x3}^2 + r_{x3}^2 r_{x1}^2 + \nonumber \\
& + r_{y1}^2 r_{y2}^2 + r_{y2}^2 r_{y3}^2 + r_{y3}^2 r_{y1}^2 + r_{x1}^2 \left(r_{y2}^2 + r_{y3}^2\right) + r_{x2}^2 \left(r_{y3}^2 + r_{y1}^2\right) + \nonumber \\
& + r_{x3}^2 \left(r_{y1}^2 + r_{y2}^2\right) + \frac{\mu^2}{9}\left(r_{x1}^2 + r_{x2}^2 + r_{x3}^2\right) + \frac{\mu^2}{36}\left(r_{y1}^2 + r_{y2}^2 + r_{y3}^2\right) - \nonumber \\
& - 2\mu\,\epsilon_{ijk}\tilde{x}_{1i}\tilde{x}_{2j}\tilde{x}_{3k} \Bigg]. \qquad \label{PotentialSpherical1}
\end{IEEEeqnarray}
\indent The above potential \eqref{PotentialSpherical1} contains four different kinds of terms, either repulsive or attractive: (1) kinetic/angular momentum terms (repulsive), (2) quartic interaction terms (attractive), (3) mass terms (attractive), and (4) a cubic Myers flux term (repulsive). The last two kinds of terms (i.e.\ the mass terms (3) and the Myers term (4)) are $\mu$-dependent and so they drop out in the $\mu \rightarrow 0$ limit (flat space) that was studied in \cite{AxenidesFloratos07}. In both cases (either $\mu = 0$ or $\mu \neq 0$), it is the equilibration of attractive and repulsive forces that determines the extrema of the potential. The two extra repulsive/attractive terms for $\mu \neq 0$ (induced by the plane-wave background) increase the complexity of the resulting dynamical system, as it will become apparent below. \\
\indent There are three ways to distribute the $\mathfrak{so}\left(3\right)$ coordinates $x_i$ ($i = 1,2,3$) into the three groups that are specified by the units $e_i$ in \eqref{Epsilon1}, so that we can generally distinguish three main types of membrane configurations. The first two of them (labelled types I and II below) describe rotating membranes (tops) that are point-like (collapsed) in one or two $\mathfrak{so}\left(3\right)$ directions and have a vanishing Myers flux term. The third type (III) is probably the most interesting one as it contains all four kinds of repulsive and attractive terms that we described above and extends into the full geometric background of $\mathfrak{so}\left(3\right) \times \mathfrak{so}\left(6\right)$. Let us now introduce these three types of configurations.
\subsection[Type I]{Type I: $q_1 = 3$, $q_2 = q_3 = 0$}
For $q_1 = 3$, $q_2 = q_3 = 0$ we have
\begin{IEEEeqnarray}{ll}
r_{x} \equiv r_{x1}, \quad r_{x2} = r_{x3} = 0 \quad \& \quad \ell_x \equiv \ell_{x1}, \quad \ell_{x2} = \ell_{x3} = 0
\end{IEEEeqnarray}
and the flux term vanishes. The effective potential \eqref{PotentialSpherical1} of the membrane becomes:
\begin{IEEEeqnarray}{ll}
V_{\text{eff}} = \frac{2\pi T}{3} \Bigg[&\frac{\ell_{x}^2}{r_{x}^2} + \frac{\ell_{y1}^2}{r_{y1}^2} + \frac{\ell_{y2}^2}{r_{y2}^2} + \frac{\ell_{y3}^2}{r_{y3}^2} + r_{y1}^2 r_{y2}^2 + r_{y2}^2 r_{y3}^2 + r_{y3}^2 r_{y1}^2 + r_{x}^2 \left(r_{y2}^2 + r_{y3}^2\right) \nonumber \\
& + \frac{\mu^2r_{x}^2}{9} + \frac{\mu^2}{36}\left(r_{y1}^2 + r_{y2}^2 + r_{y3}^2\right)\Bigg]. \qquad \label{PotentialSpherical2}
\end{IEEEeqnarray}
Apart from the completely symmetric (single-radius) configuration $r = r_x = r_{y1} = r_{y2} = r_{y3}$, $\ell = \ell_x = \ell_{y1} = \ell_{y2} = \ell_{y3}$, the radii and the momenta of the effective potential \eqref{PotentialSpherical2} may be grouped into 5 different axially symmetric (2-radii) configurations and 4 more configurations with 3 different radii. Each of these potentials possesses a local minimum that corresponds to a stationary top solution with time-independent radius and nonzero total angular momentum. There are no static solutions (i.e.\ having constant radius and zero angular momentum) in this case.
\subsection[Type II]{Type II: $q_1 = 2$, $q_2 = 1$, $q_3 = 0$}
For $q_1 = 2$, $q_2 = 1$ and $q_3 = 0$,
\begin{IEEEeqnarray}{ll}
r_{x3} = 0 \quad \& \quad \quad \ell_{x2} = \ell_{x3} = 0
\end{IEEEeqnarray}
and the flux term vanishes again. The effective potential \eqref{PotentialSpherical1} becomes:
\begin{IEEEeqnarray}{ll}
V_{\text{eff}} = \frac{2\pi T}{3} \Bigg[&\frac{\ell_{x1}^2}{r_{x1}^2} + \frac{\ell_{y1}^2}{r_{y1}^2} + \frac{\ell_{y2}^2}{r_{y2}^2} + \frac{\ell_{y3}^2}{r_{y3}^2} + r_{x1}^2 r_{x2}^2 + r_{y1}^2 r_{y2}^2 + r_{y2}^2 r_{y3}^2 + r_{y3}^2 r_{y1}^2 + \nonumber \\
& + r_{x1}^2 \left(r_{y2}^2 + r_{y3}^2\right) + r_{x2}^2 \left(r_{y3}^2 + r_{y1}^2\right) + \frac{\mu^2}{9}\left(r_{x1}^2 + r_{x2}^2\right) + \nonumber \\
& + \frac{\mu^2}{36}\left(r_{y1}^2 + r_{y2}^2 + r_{y3}^2\right) \Bigg]. \qquad \label{PotentialSpherical3}
\end{IEEEeqnarray}
Although again this case does not lead to any static configuration (with constant radius and zero angular momentum), we may construct one single-radius ($r = r_{x1} = r_{x2} = r_{y1} = r_{y2} = r_{y3}$, $\ell = \ell_{x1} = \ell_{y1} = \ell_{y2} = \ell_{y3}$) solution, 13 axially symmetric (2-radii) tops and 21 tops with 3 different radii. \\
\indent For example let us consider a type II configuration with all the $\mathfrak{so}\left(6\right)$ variables set equal to zero:
\begin{IEEEeqnarray}{ll}
x_1 = x\left(\tau\right)\cdot e_1, \quad x_2 = y\left(\tau\right)\cdot e_1, \quad x_3 = z\left(\tau\right)\cdot e_2, \quad y_i = 0, \quad i = 1,\ldots,6, \qquad \label{AnsatzExample}
\end{IEEEeqnarray}
where the time-dependent part has the form \eqref{Ansatz1}. In this case the effective potential \eqref{PotentialSpherical3} becomes:
\begin{IEEEeqnarray}{c}
V_{\text{eff}} = \frac{2\pi T}{3} \Bigg[\frac{\ell^2}{x^2 + y^2} + \left(x^2 + y^2\right) z^2 + \frac{\mu^2}{9}\left(x^2 + y^2 + z^2\right)\Bigg],
\end{IEEEeqnarray}
after setting $\ell_{x1} = \ell$ for simplicity. The corresponding extremisation condition $\nabla V_{\text{eff}} = 0$ implies
\begin{IEEEeqnarray}{c}
x \, z^2 + \frac{\mu^2 x}{9} - \frac{x \, \ell^2}{\left(x^2+y^2\right)^2} = y \, z^2 + \frac{\mu^2 y}{9} - \frac{y \, \ell^2}{\left(x^2+y^2\right)^2} = z\left(x^2 + y^2\right) + \frac{\mu^2 z}{9} = 0, \nonumber
\end{IEEEeqnarray}
which is solved by
\begin{IEEEeqnarray}{c}
x^2 + y^2 = \frac{3\ell}{\mu} \quad \& \quad z = 0. \label{ExampleSolution1}
\end{IEEEeqnarray}
Complying with \eqref{Ansatz1}, we can choose e.g.:
\begin{IEEEeqnarray}{c}
x\left(\tau\right) = \sqrt{\frac{3\ell}{\mu}}\cos\frac{\mu\,\tau}{3}, \qquad y\left(\tau\right) = \sqrt{\frac{3\ell}{\mu}}\sin\frac{\mu\,\tau}{3}, \qquad z\left(\tau\right) = 0. \label{ExampleSolution2}
\end{IEEEeqnarray}
\indent Equivalently we could have directly plugged \eqref{AnsatzExample} into the equations of motion \eqref{xEquation}--\eqref{yEquation}:
\begin{IEEEeqnarray}{l}
\ddot{x}\cdot e_1 = -x\, z^2 \cdot e_1 - \frac{\mu^2 x}{9}\cdot e_1 + \mu\,y\,z\cdot e_3 \\
\ddot{y}\cdot e_1 = -y\, z^2 \cdot e_1 - \frac{\mu^2 y}{9}\cdot e_1 + \mu\,x\,z\cdot e_3 \\
\ddot{z}\cdot e_2 = -z\left(x^2 + y^2\right)\cdot e_2 - \frac{\mu^2 z}{9}\cdot e_2.
\end{IEEEeqnarray}
It is easily seen that any solution of the type \eqref{Ansatz1} will again satisfy \eqref{ExampleSolution1}.
\subsection[Type III]{Type III: $q_1 = q_2 = q_3 = 1$}
For $q_1 = q_2 = q_3 = 1$, we write:
\begin{IEEEeqnarray}{ll}
x_1 = r_{x1} e_1, \ x_2 = r_{x2} e_2, \ x_3 = r_{x3} e_3 \quad \& \quad \ell_{x1} = \ell_{x2} = \ell_{x3} = 0. \label{TypeIII}
\end{IEEEeqnarray}
Note that $r_{x1}$, $r_{x2}$, $r_{x3}$ are not radii anymore, but coordinates. The effective potential \eqref{PotentialSpherical1} of the membrane can be written as:
\begin{IEEEeqnarray}{ll}
V_{\text{eff}} = \frac{2\pi T}{3} \Bigg[&\frac{\ell_{y1}^2}{r_{y1}^2} + \frac{\ell_{y2}^2}{r_{y2}^2} + \frac{\ell_{y3}^2}{r_{y3}^2} + r_{x1}^2 r_{x2}^2 + r_{x2}^2 r_{x3}^2 + r_{x3}^2 r_{x1}^2 + r_{y1}^2 r_{y2}^2 + r_{y2}^2 r_{y3}^2 + \nonumber \\
& + r_{y3}^2 r_{y1}^2 + r_{x1}^2 \left(r_{y2}^2 + r_{y3}^2\right) + r_{x2}^2 \left(r_{y3}^2 + r_{y1}^2\right) + r_{x3}^2 \left(r_{y1}^2 + r_{y2}^2\right) + \nonumber \\
& + \frac{\mu^2}{9}\left(r_{x1}^2 + r_{x2}^2 + r_{x3}^2\right) + \frac{\mu^2}{36}\left(r_{y1}^2 + r_{y2}^2 + r_{y3}^2\right) - 2\mu r_{x1} r_{x2} r_{x3} \Bigg]. \qquad \label{PotentialSpherical4}
\end{IEEEeqnarray}
By combining the various radii (along with the corresponding angular momenta) into groups of one, two or three, we obtain 30 different top configurations, one of which corresponds to a completely symmetric top, 9 to axially symmetric (2-radii) tops and 10 to tops that have 3 different radii.
\section[Simple type III solutions]{Simple type III solutions \label{Section:ParticularSolutions}}
\noindent The $\mathfrak{so}\left(3\right) \times \mathfrak{so}\left(3\right) \times \mathfrak{so}\left(3\right) \subset \mathfrak{so}\left(3\right) \times \mathfrak{so}\left(6\right)$ invariant ansatz
\begin{IEEEeqnarray}{ll}
x_i = \tilde{u}_i\left(\tau\right) e_i, \qquad & y_j = \tilde{v}_j\left(\tau\right) e_j, \qquad y_{j+3} = \tilde{w}_j\left(\tau\right) e_j, \qquad i,j = 1,2,3\qquad \quad \label{AnsatzSO3}
\end{IEEEeqnarray}
was studied in \cite{AxenidesFloratosLinardopoulos17a}. The ansatz \eqref{AnsatzSO3} is obviously of the form \eqref{TypeIII} (type III) and it describes rotating and pulsating membranes of spherical topology. The corresponding Hamiltonian
\begin{IEEEeqnarray}{l}
H = \frac{2\pi T}{3}\left(\tilde{p}_u^2 + \tilde{p}_v^2 + \tilde{p}_w^2\right) + U, \label{Hamiltonian}
\end{IEEEeqnarray}
is obtained by integrating out the worldvolume coordinates $\theta$ and $\phi$. The potential energy $U$ reads
\begin{IEEEeqnarray}{ll}
U = \frac{2\pi T}{3}\bigg[&\tilde{u}_1^2 \tilde{u}_2^2 + \tilde{u}_2^2 \tilde{u}_3^2 + \tilde{u}_3^2 \tilde{u}_1^2 + \tilde{r}_1^2 \tilde{r}_2^2 + \tilde{r}_2^2 \tilde{r}_3^2 + \tilde{r}_3^2 \tilde{r}_1^2 + \tilde{u}_1^2 \left(\tilde{r}_2^2 + \tilde{r}_3^2\right) + \nonumber \\
& + \tilde{u}_2^2 \left(\tilde{r}_3^2 + \tilde{r}_1^2\right) + \tilde{u}_3^2 \left(\tilde{r}_1^2 + \tilde{r}_2^2\right) + \frac{\mu^2}{9}\left(\tilde{u}_1^2 + \tilde{u}_2^2 + \tilde{u}_3^2\right) + \nonumber \\
& + \frac{\mu^2}{36}\left(\tilde{r}_1^2 + \tilde{r}_2^2 + \tilde{r}_3^2\right) - 2\mu \tilde{u}_1 \tilde{u}_2 \tilde{u}_3 \bigg], \quad \tilde{r}_j^2 \equiv \tilde{v}_j^2 + \tilde{w}_j^2, \ j = 1,2,3. \qquad \label{Potential1}
\end{IEEEeqnarray}
\indent The manifest $\mathfrak{so}\left(2\right) \times \mathfrak{so}\left(2\right) \times \mathfrak{so}\left(2\right)$ symmetry of the Hamiltonian \eqref{Hamiltonian}--\eqref{Potential1} with respect to the $\mathfrak{so}\left(6\right)$ coordinates $\tilde{v}_i$ and $\tilde{w}_i$ implies that any solution of the equations of motion preserves three $\mathfrak{so}\left(2\right)$ angular momenta $\ell_i$ ($i = 1,2,3$). The kinetic terms of the Hamiltonian \eqref{Hamiltonian} can be expressed in terms of the conserved angular momenta $\ell_i$ as
\begin{IEEEeqnarray}{l}
\tilde{p}_v^2 + \tilde{p}_w^2 = \sum_{i=1}^3\left(\dot{\tilde{r}}_i^2 + \frac{\ell_i^2}{\tilde{r}_i^2}\right),
\end{IEEEeqnarray}
leading to the effective potential
\begin{IEEEeqnarray}{ll}
V_{\text{eff}} = U + \frac{2\pi T}{3}\left(\frac{\ell_1^2}{\tilde{r}_1^2} + \frac{\ell_2^2}{\tilde{r}_2^2} + \frac{\ell_3^2}{\tilde{r}_3^2}\right). \label{EffectivePotential}
\end{IEEEeqnarray}
\subsection[The $\mathfrak{so}\left(3\right)$ symmetric membrane]{The $\mathfrak{so}\left(3\right)$ symmetric membrane}
Let us now consider the simplest possible subsystem of \eqref{AnsatzSO3} where the $\mathfrak{so}\left(6\right)$ variables $\tilde{v}_i$ and $\tilde{w}_i$ are set to zero \cite{AxenidesFloratosLinardopoulos17a}:
\begin{IEEEeqnarray}{ll}
\tilde{v}_i = \tilde{w}_i = 0, \qquad i = 1,2,3.
\end{IEEEeqnarray}
Scaling out the mass parameter $\mu$ by setting
\begin{IEEEeqnarray}{c}
t = \mu \tau, \qquad \tilde{u}_i = \mu u_i
\end{IEEEeqnarray}
leads to the form
\begin{IEEEeqnarray}{c}
V_{\text{eff}} = \frac{2\pi T \mu^4}{3}\left[u_1^2 u_2^2 + u_2^2 u_3^2 + u_1^2 u_3^2 + \frac{1}{9}\left(u_1^2 + u_2^2 + u_3^2\right) - 2 u_1 u_2 u_3\right] \label{EffectivePotentialSO3}
\end{IEEEeqnarray}
of the membrane effective potential \eqref{EffectivePotential} and the Hamilton equations of motion,
\begin{IEEEeqnarray}{ll}
\dot{u}_1 = p_1, \qquad & \dot{p}_1 = -\left[u_1\left(u_2^2 + u_3^2\right) + \frac{u_1}{9} - u_2u_3\right] \\
\dot{u}_2 = p_2, \qquad & \dot{p}_2 = -\left[u_2\left(u_3^2 + u_1^2\right) + \frac{u_2}{9} - u_3u_1\right] \\
\dot{u}_3 = p_3, \qquad & \dot{p}_3 = -\left[u_3\left(u_1^2 + u_2^2\right) + \frac{u_3}{9} - u_1u_2\right].
\end{IEEEeqnarray}
The effective potential \eqref{EffectivePotentialSO3} is a particular instance of the generalized 3-dimensional H\'{e}non-Heiles potential that was introduced in \cite{EfstathiouSadovskii04},
\begin{IEEEeqnarray}{ll}
V_{\text{HH}} = & \frac{1}{2}\left(u_1^2 + u_2^2 + u_3^2\right) + K_3 \, u_1 u_2 u_3 + K_0\left(u_1^2 + u_2^2 + u_3^2\right)^2 + \nonumber \\
& + K_4\left(u_1^4 + u_2^4 + u_3^4\right),
\end{IEEEeqnarray}
with $K_3 = -9$, $K_0 = -K_4 = 9/4$. The critical points of the effective potential \eqref{EffectivePotentialSO3} are:
\begin{IEEEeqnarray}{ll}
\textbf{u}_0 = 0, \qquad \textbf{u}_{1/6} = \frac{1}{6}\cdot\left(1, 1, 1\right), \qquad \textbf{u}_{1/3} = \frac{1}{3}\cdot\left( 1, 1, 1\right). \label{CriticalPointsSO3}
\end{IEEEeqnarray}
6 more critical points can be obtained by flipping the sign of exactly two $u_i$'s. This is consistent with the manifest tetrahedral ($T_d$) symmetry of the potential \eqref{EffectivePotentialSO3}. The extrema $\textbf{u}_0$ (point-like membrane) and $\textbf{u}_{1/3}$ (Myers dielectric sphere) are global degenerate minima of the potential while $\textbf{u}_{1/6}$ is a saddle point:
\begin{IEEEeqnarray}{c}
V_{\text{eff}}\left(0\right) = V_{\text{eff}}\left(\frac{1}{3}\right) = 0,\qquad V_{\text{eff}}\left(\frac{1}{6}\right) = \frac{2\pi T \mu^{4}}{6^4}.
\end{IEEEeqnarray}
\paragraph{Radial spectrum} \cite{AxenidesFloratosLinardopoulos17a} By radially perturbing the 9 critical points $u_i^0$ in \eqref{CriticalPointsSO3} as
\begin{IEEEeqnarray}{ll}
u_i = u_i^0 + \delta u_i\left(t\right), \qquad \delta u_i\left(t\right) = \sum_{k=1}^{6}c_k e^{i\lambda t}\xi_{ik},
\end{IEEEeqnarray}
we may confirm the above conclusion by examining the corresponding Hessian matrix. It turns out that $\textbf{u}_0$ and $\textbf{u}_{1/3}$ are global minima (positive-definite Hessian) and $\textbf{u}_{1/6}$ is a saddle point (indefinite Hessian). These results are summarized in the following table \ref{Table:RadialSpectrumSO3}. \vspace{-.7cm}
\begin{table}[h!]
\begin{center}
\begin{IEEEeqnarray}{c}
\begin{array}{|c|c|c|}
\hline && \\
\quad {\color{red}\text{critical\ point}} \quad & {\color{red}\text{eigenvalues } \lambda^2 \text{ (\#)}} & {\color{red}\text{stability}} \\[6pt]
\hline && \\
\textbf{u}_0 & \frac{1}{9} \, \left(3\right), \ \frac{1}{36} \, \left(6\right) & \text{center (S)} \\[12pt]
\textbf{u}_{1/6} & \quad -\frac{1}{18} \, \left(1\right), \ \frac{5}{18} \, \left(2\right), \ \frac{1}{12} \, \left(6\right) \quad & \quad \text{saddle point} \quad \\[12pt]
\textbf{u}_{1/3} & \frac{1}{9} \, \left(1\right), \ \frac{4}{9} \, \left(2\right), \ \frac{1}{4} \, \left(6\right) & \text{center (S)} \\[6pt]
\hline
\end{array} \nonumber
\end{IEEEeqnarray}
\end{center}
\caption{Radial spectrum of the $\mathfrak{so}\left(3\right)$ symmetric membrane. \label{Table:RadialSpectrumSO3}}
\end{table} \vspace{-1cm}
\paragraph{Angular spectrum} \cite{AxenidesFloratosLinardopoulos17b} We may also perform more general (angular/multipole) perturbations of the following form:
\begin{IEEEeqnarray}{c}
x_i\left(t\right) = x_i^0 + \delta x_i\left(t\right), \qquad i = 1, 2, 3,
\end{IEEEeqnarray}
where $\delta x_i$ is expanded in spherical harmonics $Y_{jm}\left(\theta,\phi\right)$ as
\begin{IEEEeqnarray}{c}
x_i\left(t\right) = \mu u_i\left(t\right)e_i, \quad x_i^0 \equiv \mu u_i^0 e_i, \quad \delta x_i\left(t\right) = \mu\cdot\sum_{j = 1}^{\infty}\sum_{m=-j}^{j}\eta_i^{jm}\left(t\right) Y_{jm}\left(\theta,\phi\right). \qquad
\end{IEEEeqnarray}
For the critical points $\textbf{u}_{0}$, $\textbf{u}_{1/6}$, $\textbf{u}_{1/3}$ we find the eigenvalues \cite{AxenidesFloratosLinardopoulos17a}:
\begin{IEEEeqnarray}{ll}
\textbf{u}_0: \ &\lambda_P^2 = \lambda_{\pm}^2 = \frac{1}{9}, \ \lambda_{\theta}^2 = \frac{1}{36} \\[6pt]
\textbf{u}_{1/6}: \ &\lambda_P^2 = 0, \ \lambda_+^2 = \frac{1}{36}\left(j + 1\right)\left(j + 4\right), \ \lambda_-^2 = \frac{j\left(j-3\right)}{36}, \nonumber \\
& \lambda_{\theta}^2 = \frac{1}{36}\left(j^2 + j + 1\right) \\[6pt]
\textbf{u}_{1/3}: \ &\lambda_P^2 = 0, \ \lambda_+^2 = \frac{1}{36}\left(j + 1\right)^2, \ \lambda_-^2 = \frac{j^2}{9}, \ \lambda_{\theta}^2 = \frac{1}{36}\left(2j + 1\right)^2, \qquad
\end{IEEEeqnarray}
with multiplicities $d_P = 2j+1$, $d_+ = 2j+3$, $d_- = 2j-1$ and $d_{\theta} = 6\left(2j+1\right)$, respectively. \\
\indent The critical point $\textbf{u}_0$ (point-like membrane) is obviously stable. $\textbf{u}_{1/3}$ has a zero mode of degeneracy $2d_P$ while all its other eigenvalues are stable for $j = 1,2,\ldots$ $\textbf{u}_{1/6}$ has one $2d_P$-degenerate zero mode for every $j$ and a 10-fold degenerate zero mode for $j=3$. It is unstable for $j=1$ (2-fold degenerate) and $j=2$ (6-fold degenerate). The above results were first obtained by \cite{DasguptaJabbariRaamsdonk02} from the matrix model. In the flat-space limit ($\mu \rightarrow 0$), we recover the results of \cite{AxenidesFloratosPerivolaropoulos00, AxenidesFloratosPerivolaropoulos01}.
\subsection[The $\mathfrak{so}\left(3\right) \times \mathfrak{so}\left(6\right)$ symmetric membrane]{The $\mathfrak{so}\left(3\right) \times \mathfrak{so}\left(3\right) \times \mathfrak{so}\left(3\right)$ symmetric membrane}
Similar perturbative analyses can be carried out in the $\mathfrak{so}\left(3\right) \times \mathfrak{so}\left(6\right)$ sector. A solution of the corresponding equations of motion is given by
\begin{IEEEeqnarray}{c}
u_i^0 = u_0, \quad v_j^0\left(t\right) = v_0 \cos\left(\omega t + \varphi_j\right), \quad w_j^0\left(t\right) \equiv v_{j+3}^0\left(t\right) = v_0 \sin\left(\omega t + \varphi_k\right), \qquad \
\end{IEEEeqnarray}
where $(u_0,v_0)$ are the critical points of the axially symmetric potential
\begin{IEEEeqnarray}{ll}
V \equiv \frac{V_{\text{eff}}}{2 \pi T \mu^4} = u^4 &+ 2 u^2 v^2 + v^4 + \frac{u^2}{9} + \frac{v^2}{36} - \frac{2 u^3}{3} + \frac{\ell ^2}{v^2}
\end{IEEEeqnarray}
and $\ell \mu^3 \equiv \ell_{1} = \ell_{2} = \ell_{3}$. It can be proven that the critical points $(u_0,v_0)$ always lie within the interval:
\begin{IEEEeqnarray}{ll}
\frac{1}{6} \leq u_0 \leq \frac{1}{3} \qquad \& \qquad 0 \leq v_0 \leq \frac{1}{12}. \label{IntervalSO3xSO6}
\end{IEEEeqnarray}
\paragraph{Radial spectrum} \cite{AxenidesFloratosLinardopoulos17a} To obtain the radial spectrum we set
\begin{IEEEeqnarray}{c}
u_i = u_i^0 + \delta u_i\left(t\right), \qquad v_i = v_i^0\left(t\right) + \delta v_i'\left(t\right), \qquad w_i = w_i^0\left(t\right) + \delta w_i'\left(t\right), \qquad
\end{IEEEeqnarray}
finding six zero eigenvalues and four nonzero ones (quadruply and doubly degenerate):
\begin{IEEEeqnarray}{l}
\lambda_{1\pm}^2 = \frac{5u_0}{2} - \frac{1}{9} \pm \sqrt{\frac{1}{9^2} - \frac{u_0}{9} - \frac{5u_0^2}{12} + 4 u_0^3} \label{RadialEigenvalues1} \\
\lambda_{2\pm}^2 = \frac{5u_0}{2} - \frac{5}{18} \pm \sqrt{\frac{5^2}{18^2} - \frac{35u_0}{18} + \frac{163u_0^2}{12} - 20u_0^3}. \label{RadialEigenvalues2}
\end{IEEEeqnarray}
The plots of these eigenvalues can be found in the following figure \ref{Figure:RadialSpectrumSO3xSO6}.
\vspace{-.3cm}\begin{figure}[h!]
\begin{center}
\includegraphics[scale=0.4]{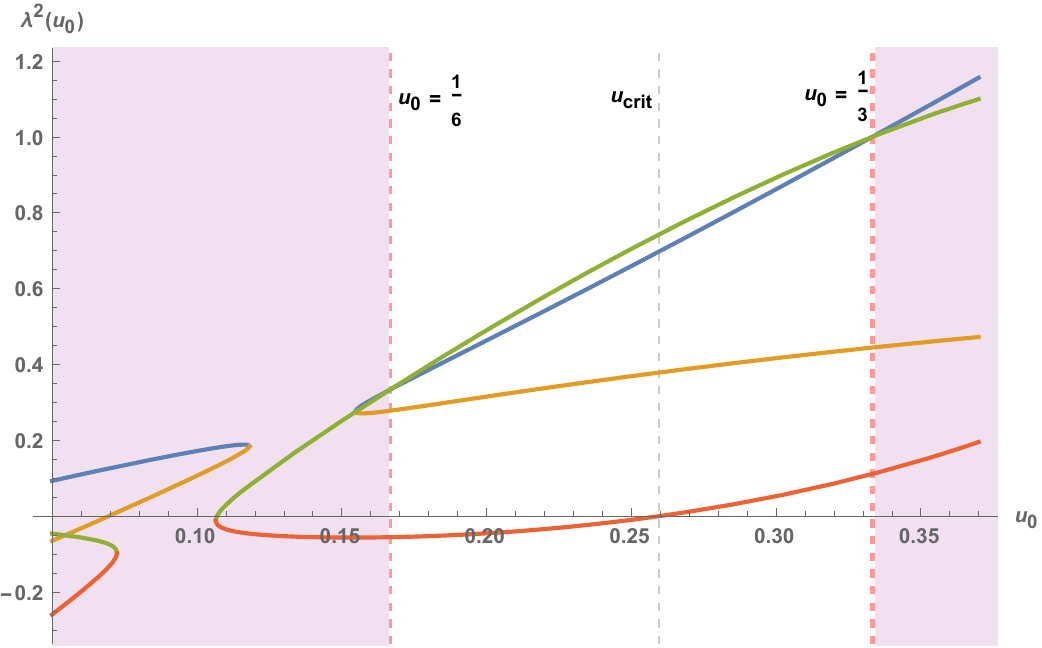}
\end{center}
\caption{Radial spectrum of the $\mathfrak{so}\left(3\right) \times \mathfrak{so}\left(6\right)$ symmetric membrane. \label{Figure:RadialSpectrumSO3xSO6}}
\end{figure} \vspace{-1cm}
\paragraph{Angular spectrum} \cite{AxenidesFloratosLinardopoulos17b} Going further, we again set out to perform angular/multipole perturbations of the form:
\begin{IEEEeqnarray}{c}
x_i = x_i^0 + \delta x_i, \quad i = 1, 2, 3 \qquad \& \qquad y_k = y_k^0 + \delta y_k, \quad k = 1,\ldots,6, \qquad
\end{IEEEeqnarray}
where the $\delta x_i, \ \delta y_k$ are expanded around the classical solution,
\begin{IEEEeqnarray}{ll}
x_i^0 = \mu u_0 e_i, \quad i = 1, 2, 3, \qquad & y_i^0 = \mu v_i^0\left(t\right) e_1, \quad i = 1,2 \\
& y_k^0 = \mu v_k^0\left(t\right) e_2, \quad k = 3,4 \\
& y_l^0 = \mu v_l^0\left(t\right) e_3, \quad l = 5,6,
\end{IEEEeqnarray}
in spherical harmonics $Y_{jm}\left(\theta,\phi\right)$:
\begin{IEEEeqnarray}{ll}
\delta x_i = \mu\cdot\sum_{j,m}\eta_i^{jm}\left(\tau\right) Y_{jm}\left(\theta,\phi\right), \qquad
&\delta y_k = \mu\cdot\sum_{j,m}\epsilon_k^{jm}\left(\tau\right) Y_{jm}\left(\theta,\phi\right) \\
&\delta y_l = \mu\cdot\sum_{j,m}\zeta_l^{jm}\left(\tau\right) Y_{jm}\left(\theta,\phi\right), \qquad
\end{IEEEeqnarray}
for $i = 1,2,3$, $k = 1,3,5$, $l = 2,4,6$. We find that one of the eigenvalues always vanishes, two others are given by the following analytic expression
\begin{IEEEeqnarray}{ll}
\lambda_{P}^2 = \frac{1}{2}\left(j^2 + j + 2\right)u_0 &- \frac{1}{18}\bigg(1 + j\left(j + 1\right) \pm \nonumber \\
& \hspace{-1.5cm} \pm 3\sqrt{144\left(j^2 + j - 2\right)u_0^3 - 12\left(j^2 + j - 14\right)u_0^2 - 24u_0 + 1}\bigg), \qquad
\end{IEEEeqnarray}
while 6 more eigenvalues $\lambda_{\pm}$ are also known in closed forms but are too complicated to be included here. The corresponding multiplicities of the eigenvalues are $d_P = 2j+1$, $d_+ = 2j+3$, $d_- = 2j-1$. For $j = 1$ four eigenvalues vanish, while two others coincide with the eigenvalues \eqref{RadialEigenvalues1}--\eqref{RadialEigenvalues2} that were found from radial perturbations:
\begin{IEEEeqnarray}{ll}
\lambda_P^2 = 4u_0 + \frac{1}{3}, \qquad
&\lambda_{+}^2 = \frac{5u_0}{2} - \frac{1}{9} \pm \sqrt{\frac{1}{9^2} - \frac{u_0}{9} - \frac{5u_0^2}{12} + 4 u_0^3} \\
&\lambda_{-}^2 = \frac{5u_0}{2} - \frac{5}{18} \pm \sqrt{\frac{5^2}{18^2} - \frac{35u_0}{18} + \frac{163u_0^2}{12} - 20u_0^3}. \qquad
\end{IEEEeqnarray}
For $j = 2$ there's one zero eigenvalue while $\lambda_P > 0$. We can also plot the $j = 2$ eigenvalues of $\lambda_{\pm}$ (figure \ref{Figure:AngularSpectrumSO3xSO6}):
\begin{figure}[h!]
\begin{center}
\includegraphics[scale=0.3]{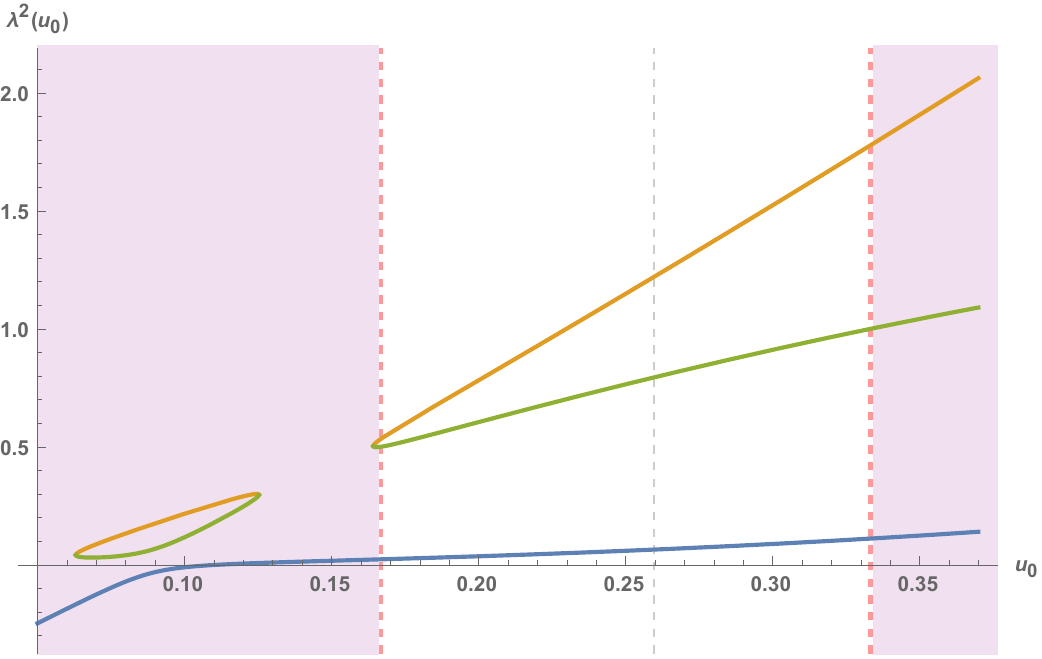} \quad \includegraphics[scale=0.3]{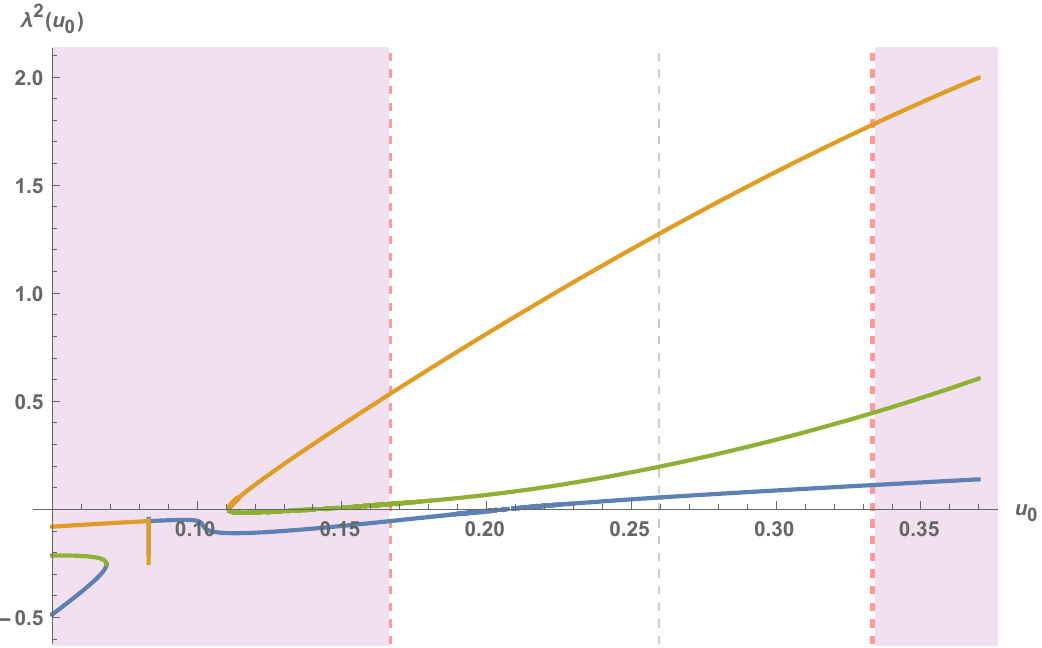}
\end{center}
\caption{$\lambda_{\pm}^2$ for $j = 2$ as a function of $u_0$. \label{Figure:AngularSpectrumSO3xSO6}}
\end{figure} 
\begin{itemize}
\item The squared nonzero $j = 1$ eigenvalues are all positive/stable in the interval \eqref{IntervalSO3xSO6}, except $\lambda_{-(-)}^2$ which is positive/stable only for $u_{\text{crit}} < u_0 < 1/3$, where $u_{\text{crit}} \equiv \frac{1}{60} \left(11+\sqrt{21}\right)$.\pagebreak
\item For $j = 2$, the $\lambda_P$, $\lambda_+$ and one of the $\lambda_-$ squared eigenvalues are positive/stable in the interval \eqref{IntervalSO3xSO6}. The remaining $\lambda_-^2$ eigenvalue is negative/unstable in the interval $\frac{1}{6} \leq u_0 \leq 0.207245 < u_{\text{crit}}$. \\
\item For $j \geq 3$ all the squared eigenvalues are non-negative inside the interval \eqref{IntervalSO3xSO6} and so the system is stable .
\end{itemize}
Here's a summary of the angular/multipole spectrum (table \ref{Table:AngularSpectrumSO3xSO6}):
\vspace{-.6cm}\begin{table}[h!]
\begin{center}
\begin{eqnarray}
\begin{array}{|c|c|c|c|c|}
\hline &&&& \\
\ {\color{red}\text{eigenvalues}} \ & {\color{red}j=1} & {\color{red}j=2} & {\color{red}j \geq 3} & {\color{red}\text{degeneracy}} \\[6pt]
\hline &&&& \\
\lambda_P^2 & 0,0,+ & 0,+,+ & 0,+,+ & d_P = 2j+1 \\[12pt]
\lambda_+^2 & 0,+,+ & +,+,+ & +,+,+ & d_+ = 2j+3 \\[12pt]
\lambda_-^2 & 0,+,\{0,\pm\} & +,+,\{0,\pm\} & \ +,+,+ \ & \ d_- = 2j-1 \ \\[3pt]
& \ \Big(\begin{array}{c} \text{positive for} \\ u_0 > u_{\text{crit}} \end{array}\Big) \ & \ \Big(\begin{array}{c} \text{positive for} \\ u_0 > 0.207245 \end{array}\Big) \ && \\[6pt]
\hline
\end{array} \nonumber
\end{eqnarray}
\end{center}
\caption{Angular spectrum of the $\mathfrak{so}\left(3\right) \times \mathfrak{so}\left(6\right)$ symmetric membrane. \label{Table:AngularSpectrumSO3xSO6}}
\end{table} \vspace{-1cm}
\paragraph{Higher-order perturbations} \cite{AxenidesFloratosKatsinisLinardopoulos21} Beyond linearized perturbation theory (always inside the interval \eqref{IntervalSO3xSO6}), we anticipate a cascade of instabilities that originates from the $j = 1,2$ multipoles and propagates towards all higher modes ($j = 3,4, \ldots$). The perturbative expansion becomes
\begin{IEEEeqnarray}{l}
x_i = \sum_{n = 0}^{\infty}\varepsilon^n\delta x_i^n = x_i^0 + \sum_{n = 1}^{\infty}\varepsilon^n\delta x_i^n, \quad i = 1, 2, 3 \\
y_i = \sum_{n = 0}^{\infty}\varepsilon^n\delta y_i^n = y_i^0 + \sum_{n = 1}^{\infty}\varepsilon^n\delta y_i^n, \quad i = 1,\ldots,6.
\end{IEEEeqnarray}
It follows that any given mode $j$ at any given order $n$ in perturbation theory couples to all the modes of the previous orders $1,\ldots, n-1$ through an effective forcing term that emerges in the corresponding system of fluctuation equations. The perturbations are expanded in spherical harmonics as
\begin{IEEEeqnarray}{lll}
\delta x_i^n = \mu\cdot\sum_{j,m}\eta_i^{njm}\left(\tau\right) Y_{jm}\left(\theta,\phi\right), \qquad &\eta_i^{njm}\left(0\right) = 0, \qquad &i = 1,2,3 \\
\delta y_i^n = \mu\cdot\sum_{j,m}\theta_i^{njm}\left(\tau\right) Y_{jm}\left(\theta,\phi\right), \qquad &\theta_i^{njm}\left(0\right) = 0, \qquad &i = 1,\ldots,6.
\end{IEEEeqnarray}
For example it can be shown that the $(n = 1, \, j = 1,2)$ instabilities we found above couple to every mode ($j = 1, 2,\ldots$) of the second order ($n = 2$) in perturbation theory.
\section[Acknowledgements]{Acknowledgements}
The authors would like to thank Christos Skiadas for the invitation to participate to the $13^{th}$ CHAOS conference. The present research is funded in the context of the project "Chaotic dynamics and black holes in BMN theory" E-12386 (MIS 5047794) under the call for proposals "Supporting researchers with an emphasis on young researchers---Cycle B" (EDULLL 103). The project is co-financed by Greece and the European Union (European Social Fund---ESF) by the Operational Programme Human Resources Development, Education and Lifelong Learning 2014-2020.

\bibliographystyle{JHEP}
\bibliography{D:/Documents/1.Research/3.Athenes/2.TeXFiles/Bibliography/JHEP_Bibliography,D:/Documents/1.Research/3.Athenes/2.TeXFiles/Bibliography/Math_Bibliography}
\end{document}